\documentclass[%
 aip,
 jmp,%
 amsmath,amssymb,
preprint,%
]{revtex4-1}

\usepackage{graphicx}
\usepackage{dcolumn}
\usepackage{bm}
\usepackage{textcomp,gensymb}
\usepackage{caption,subcaption}

\begin{document}


\title[]{Numerical modeling of ion transport in a ESI-MS system}

\author{Natalia Gimelshein}
\affiliation{Gimel Inc, CA 91020}
\author{Sergey Gimelshein}
\affiliation{University of Southern California, Los Angeles, CA 91020}
\author{Taylor Lilly}
\affiliation{University of Colorado Colorado Springs, Colorado
  Springs, CO 80918}
\author{Eugene Moskovets}
\affiliation{MassTech, Columbia, MD 21046}


\begin{abstract}
Gas and ion transport in the capillary-skimmer subatmospheric interface
of a mass spectrometer, which is typically utilized to separate
unevaporated micro-droplets from ions, was studied numerically using a
two-step approach spanning multiple gas dynamic regimes. The gas flow
in the heated capillary and in the interface was determined by solving
numerically the Navier-Stokes equation. The capillary-to-skimmer
gas/ion flow was modeled through the solution of the full Boltzmann
equation with a force term. The force term, together with calculated
aerodynamic drag, determined the ion motion in the gap between the
capillary and skimmer. The three-dimensional modeling of the impact of
the voltage applied to the Einzel lens on the transmission of
doubly-charged peptides ions through the skimmer orifice was compared
with experimental data obtained in the companion study. Good agreement
between measured and computed signals was observed. The numerical
results indicate that as many as 75\% ions that exit from the capillary
can be lost on the conical surface of the skimmer or capillary outer
surface due to the electrostatic force and plume divergence.
\end{abstract}

\maketitle


\newcommand*{\citen}[1]{%
  \begingroup
    \romannumeral-`\x 
    \setcitestyle{numbers}%
    [\cite{#1}]%
  \endgroup   
}

\section{Introduction}

The development of electrospray ionization (ESI) mass spectrometry
(MS) 25 years ago has opened an opportunity to characterize substances
of broad nature\cite{Smith,Kebarle,Fenn}, most notably proteins.  A
major deficiency of operation of mass analyzers utilizing atmospheric
ion sources is that only a small fraction of molecular ions generated
from charged microdroplets reaches the ion detector. A significant
loss of analytes takes place during transport of the ions through the
MS inlet system\cite{Lin}, which is often a long heated capillary with
a 400-600~$\mu$m i.d. followed by a subatmospheric chamber with a
skimmer; the latter separates the low vacuum from the high vacuum
sections of the MS system.  A two-fold increase in the ion
transmission from a nano-ESI source to the mass analyzer was
demonstrated on a flared inlet capillary \cite{Wu}.  More substantial
increase in the ion transmission from an ESI source to the mass
analyzer was achieved using ion funnels \cite{Kelly}. In general, more
efficient operation of ESI sources coupled with a heated inlet
capillary and subamospheric/low vacuum ion transport system downstream
of the inlet has a tremendous impact on the field of biological mass
spectrometry. Such an operation may be achieved by reducing the
unspecific chemical noise, the loss of analyte ions on capillary walls
or orifices providing differential pumping, and the in-source
fragmentation of fragile ions, which determines the detection limit in
a particular mass analyzer and the time needed for individual MS and
MS/MS analyses.

Considerable progress in reducing ion loss in atmospheric interfaces
is visibly hampered by a lack of comprehensive and accurate system
modeling, both at the microscopic physical and fluid dynamic levels.
Ideally, this comprehensive modeling would comprise the accurate
simulation of droplet evolution (fission and evaporation), production
of ions from droplets, and ion loss inside the heated capillary and on
the MS inlet elements downstream. At present, there is no
self-consistent and complete model capable of numerical simulation of
such an evolution.  Instead, researchers study and analyze different
aspects of that evolution, separated in physical and spatial terms.
For example, droplet evaporation and fission is usually separated from
gas/ion transport, and gas flow in the inlet capillary is separated
from the flow in the subatmospheric sections.  There are several
reasons for such simplifications. The most important is the change in
flow regimes from viscous to molecular over a short spatial scale,
associated with the change in gas density from atmospheric levels at
the inlet to microTorr levels in the high vacuum section of the MS.
There is also a dramatic change in time scale - from slowly changing
flow in the long inlet capillary to rapidly propagating supersonic
expansion flow downstream of the capillary.

Yet another complexity for comprehensive modeling of ESI-MS
droplet/ion evolution is related to droplet-to-ion transformation. The
transport of micron size droplets of solvent and ions in carrier gas
flow is governed primarily by drag force from the carrier gas and, in
addition, by electrostatic force from applied fields and space charge.
While the droplet transport often may be uncoupled from gas transport,
the difficulty of modeling the evolution of micro- and nanodroplets
stems from the lack of detailed microscopic information (mostly due to
experimental challenges) on droplet evaporation and
fission\cite{Hogan}. Lagrangian particle tracking may be used for
modeling droplet evolution \cite{Longest2004}, and a CFD solver based
on the solution of Navier-Stokes equations for droplet
transport\cite{Longest2007,wup}.  Depending on the selected capillary
temperature and, importantly, the temperature gradient\cite{Eugene},
the formation of analyte ions from nanodroplets usually occurs not in
the capillary head but inside a longer section of the capillary, so
the transport of these ions through the capillary should take into
consideration gas densities that can be significantly lower than
atmospheric.  Knudsen numbers calculated from the characteristic flow
dimensions, such as capillary i.d. or skimmer diameter, typically are
on the order 0.01 to 0.1 for flows expanding through the capillary,
and 0.1 to 10 in the ion transport systems kept at typical pressures
of 1 Torr (ion funnels) or 0.01 Torr (RF multipoles). The flows at
such Knudsen numbers are in a regime transitional from continuum to
free molecular.  Neither continuum formulation, such as that based on
the Navier-Stokes equations, or free molecular, such as test particle
methods, are applicable to compute gas flows at such conditions.
Numerical analysis of such flows requires a kinetic approach to be
used and thus the consideration of the Boltzmann equation, the
fundamental equation of the kinetic theory, that properly accounts for
flow non-equilibrium.

The numerical complexity and high computational cost of a
three-dimensional solution of the Boltzmann equation that inherently
involves development and application of highly scalable parallel
algorithms are the obvious reasons for its very rare application to
ESI-MS ion transportation problems. Since deterministic approaches to
the solution of the Boltzmann equation, such as its numerical
integration, are prohibitively expensive from the computer time stand
point, the only feasible way to solve it is statistical, based on the
direct simulation Monte Carlo (DSMC) method \cite{DSMC}.  The known
applications of the DSMC method for gas/droplet/ion transport in
ICP-MS are Ref.~\citen{debbie} where evaporation and coalescence
affected droplet evolution was studied, Ref.~\citen{greek} where the
carrier gas flow inside an ion funnel was examined, and
Ref.~\citen{ouyang} where carrier gas flow from a capillary into a
skimmer was modeled. Note that Ref.~\citen{greek} represented, in
fact, a three-step approach where the DSMC inflow boundary conditions
at the inlet capillary exit were uniform values approximated from
Navier-Stokes simulations and the DSMC fields were in turn used for
successive particle tracking with a SIMION package.  In
Ref.~\citen{ouyang}, a two-dimensional setup was used with a uniform
capillary exit approximation; the calculations of ion trajectories
were not presented.

This work is the first step toward a comprehensive model of ion
transport in ESI-MS systems, which includes droplet evaporation,
fission, and transport, along with the formation, transport, and loss
of ionized analytes and solvent clusters (residual nanodroplets) in a
wide range of $m/z$. A pragmatic approach to satisfying this goal is
to first simulate the backbone of the process, the multi-regime gas
dynamic transfer from atmosphere to high vacuum.  Once established, a
robust capability for modeling gas dynamics in the system will act as
the foundation on which to add each of the additional pieces to the
comprehensive model.  For instance, accurate thermal and pressure data
is required as inputs to droplet evaporation. Local gas flow velocity,
density, and temperature, as well as collisional cross sections, are
required for compute ion transport in subatmospheric and low-vacuum
sections of mass spectrometers. Thus, this work focuses on
establishing an accurate gas dynamic backbone as the first step to a
comprehensive ESI-MS numerical model.  Consequently in this work, ions
are not considered until the DSMC portion of the simulation and then
introduced near the capillary exit with properties corresponding to
the local carrier gas macroparameters. This assumption will be
addressed later.  While predominately gas dynamic, this limited
inclusion of ions allows the simulation to be compared directly with
experiment while changing only one key parameter - Einzel lens
potential.  This will be used as the characteristic of interest in
assessing the accuracy of the gas dynamics simulations in
subatmospheric interfaces and the validity of this model as the
foundation for future additions and refinements.

The DSMC method is used in this work to model the ion transport in a
carrier gas flow expanding through a long capillary into a
low-pressure background gas before passing through a skimmer into the
hexapole chamber. The main objective of the work is to analyze the
efficiency of the traditional capillary/Einzel lens/off-center skimmer
approach, currently used in portable mass
spectrometers\cite{Misharin}, and study the effect of changing voltage
at the Einzel (tube) lens. The companion experimental study provides
not only the actual geometric and flow properties of interest, but
also solid ground for model validation.  The off-center location of
the skimmer does not allow one to use the simplifying assumption of an
axially symmetric flow utilized in previous studies\cite{debbie,greek}.
Thus, full three-dimensional modeling of the ion-gas mixture is
performed here. A CFD software package, CFD++, was utilized for
modeling the capillary flow that provided the hand-off starting
surface for the DSMC simulations, while a multi-physics solver COMSOL
was used to calculate the electric field and related force term in the
Boltzmann equation.

There are several issues considered in this work that were omitted in
previous research. First, and most notably, this is the first truly
three-dimensional simulation that uses a kinetic approach to model the
subatmospheric and low vacuum sections of the MS system.  The
application of a kinetic approach based on the solution of the
Boltzmann equation provides accurate description of the flow as it
takes into account its microscopic nature at the level of the velocity
distribution functions, thus resolving flow non-equilibrium.  Flow
three-dimensionality is critical as it allows one to consider the
widely used off-axis setup that greatly reduces the contamination of
the MS system downstream with residual droplets. Second, the
simulation includes fully coupled ion-carrier gas flow in the low
vacuum section.  Third, the numerical modeling presented in this work
takes into consideration the flow inside a capillary as a precursor to
the flow in the low vacuum section, thus avoiding a uniform inflow
approximation that may result in significant accuracy
loss\cite{sidejet}.

\section{Numerical Approach and Flow Conditions}

In the physical system, micron-size solvent droplets are
formed at the tip of a Taylor cone formed in the ESI source, and move
through atmosphere into the inlet capillary. The capillary has an
inner diameter of 0.5~mm and a total length of 100~mm. In order to
provide high-temperature environment needed for droplet evaporation
and subsequent ion formation, the capillary is heated to 227~C
(500~K). The droplet fission and evaporation in the capillary produces
ions mostly by charge residue\cite{Dole}, but also by ion
evaporation\cite{Iribarne} mechanisms.  Then, the mixture of ions and
carrier gas molecules expands out of the inlet capillary into a
chamber where gas pressure is kept at 1~Torr.  As the ions travel in
the plume, they are focused by the electrostatic force produced by the
potential difference between the Einzel lens, capillary, and skimmer.
Some of them then move through the 0.6~mm diameter skimmer into the
hexapole chamber kept at 0.01~Torr.

\begin{figure}[htb]
  \begin{subfigure}[b]{\linewidth}
    \centering  
    \includegraphics[width=13cm]{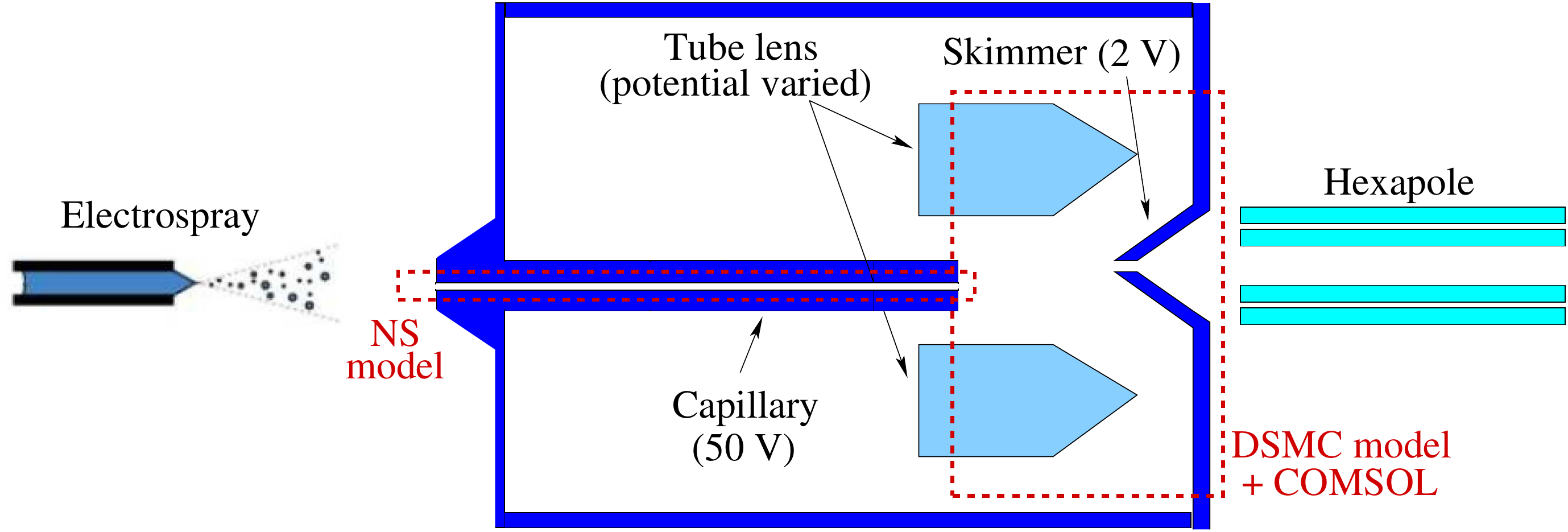}
    \caption{}\label{scheme}
  \end{subfigure}\\%
  \begin{subfigure}[b]{\linewidth}
    \centering
    \includegraphics[width=14cm]{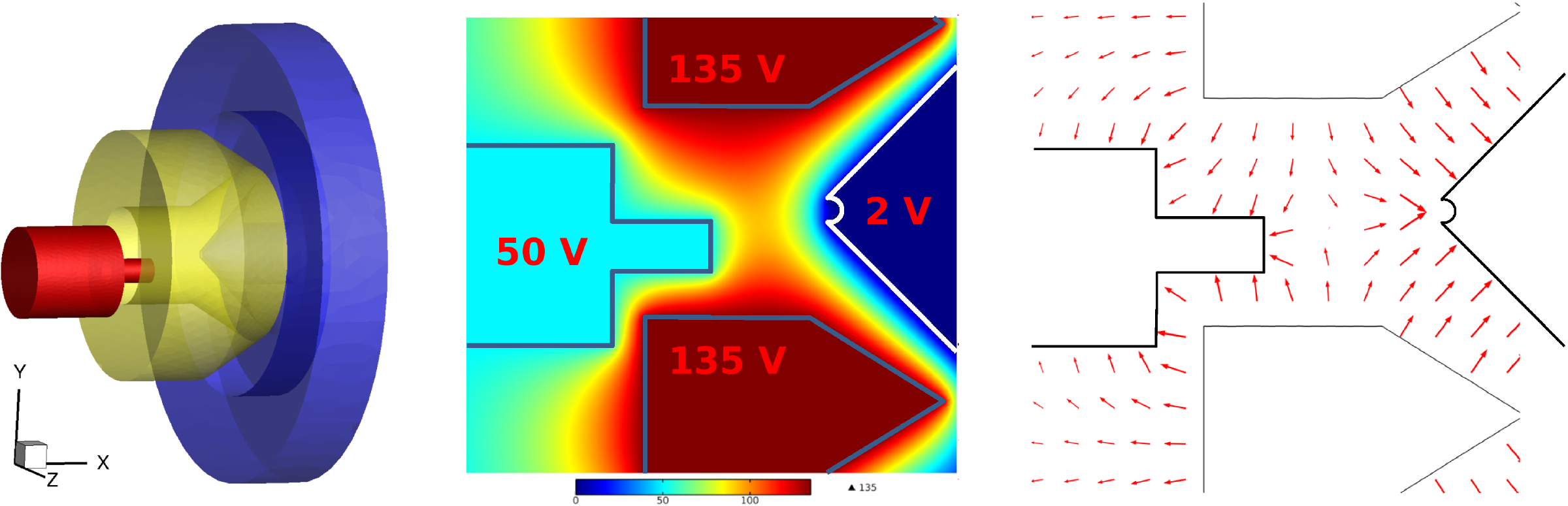}
    \caption{}\label{COMSOL}
  \end{subfigure}%
  \caption{(a) Schematics of ESI-MS setup and (b) numerical modeling
    approaches and geometries used in COMSOL and SMILE. In (b), left,
    are the models for the capillary (red), Einzel lens (yellow), and
    skimmer (blue); in (b), center, is a two-dimensional slice of
    electric potential; in (b), right, is the electric field vectors.
    Note that the lens and the skimmer are shifted 1 mm in the upward
    direction.}
\end{figure}

A partially simplified geometry of the skimmer in the first
subatmospheric chamber and the Einzel lens in the inlet of the mass
spectrometer was used in computations. The schematics of the
considered setup is plotted in Fig.~\ref{scheme} in a plane aligned
with the inlet capillary axis and the skimmer axis.  This geometry has
been utilized in earlier models of quadrupole ion traps (Finnigan LCQ,
ThermoFisher, San Jose) and in several mass spectrometers equipped
with atmospheric interfaces built by MassTech Inc. Note that the
center of the skimmer is shifted off the capillary axis to avoid
contamination of downstream chambers by droplets not fully evaporated
in the heated inlet capillary.

Due to the five order of magnitude pressure drop from the ESI source
to the hexapole chamber, subsonic flow in the inlet capillary that
requires long time to reach steady state, and three-dimensionality of
the skimmer chamber setup, it is practical to separate the problem
into two steps. The first step considers the flow inside the
capillary; it starts in the ESI chamber and ends in the first
subatmospheric chamber of the MS.  The gas flow regime is continuum at
the capillary entrance and near-continuum at its exit. Thus, the
Navier-Stokes equations may be used to calculate that flow.
Additionally, the flow is axially symmetric. The computational domain
used for the first step is schematically shown in Fig.~\ref{scheme},
denoted as ``NS model''. The second step includes the flow from the
capillary to the skimmer. The flow regime is transitional, and the
DSMC method is therefore used (denoted ``DSMC model'' in
Fig.~\ref{scheme}).  It is important to mention that an accurate
splitting of the flow into an internal and external parts requires the
hand-off surface to be located in the supersonic region. The hand-off
surface serves as the inflow boundary condition for second step, with
flow properties obtained in the first step. The supersonic velocity at
that surface guarantees that gas molecules and ions from the
downstream portion of the plume do not travel upstream. It is usually
sufficient\cite{ss} to set the hand-off surface at a Mach number $M=3$
isosurface.

Modeling of the first step was performed with a computational tool
CFD++ \cite{CFD++}.  CFD++ is a flexible computational fluid dynamics
software suite developed by Metacomp Technologies for the solution of
steady and unsteady, compressible and incompressible Navier-Stokes
equations, including multi-species capability for perfect and reacting
gases. CFD++ was proven to be a robust tool for complex gas dynamic
flows, both external and internal\cite{cfd1,cfd2}. In this work, a
turbulent Reynolds Averaged Navier-Stokes capability of CFD++ is
applied. The simulations were conducted with a 2nd order in space,
Harten, Lax, van Leer, contact discontinuity Riemann approximation
algorithm. The computations use a three-equation $k-\epsilon-R_t$
turbulence model \cite{turb}. Implicit time integration was used. A
perfect gas model of air was used in this work as the impact of real
gas effects for the temperatures under consideration is negligible .
The capillary surface was assumed isothermal with a constant
temperature of 500~K. Second order slip boundary conditions were
applied at the wall. A symmetry condition is used at the capillary
axis. A stagnation pressure-temperature boundary condition was set at
the inflow boundary and a backpressure imposition was applied at the
outflow.  Note that the inflow and outflow boundaries were set
sufficiently far from the capillary walls so that the Mach number
$M\ll 1$ at the former and $M\gg 1$ at the latter. A multi-block
computational grid was used with the total number of cells over
200,000. Note that solution convergence was verified for several grids
to obtain grid independent results.

Modeling of the second step was conducted with a DSMC solver SMILE
\cite{SMILE}.  The three-dimensional parallel capability of SMILE was
used here. In the past, SMILE has been extensively used to solve the
Boltzmann equation without and with the force term, and validated in a
number of multi-dimensional problems \cite{smile1,smile2,smile3}. The
majorant frequency scheme\cite{MFS} was used to calculate
intermolecular interactions.  A three-species mixture was considered,
with the standard atmospheric composition of molecular nitrogen and
oxygen, and a trace species of ion with the initial mole fraction of
10$^{-5}$ (small enough not to impact the gas flow). A species
weighting scheme was applied in gas-ion collisions. Ion atomic mass
was 1348~Da (doubly charged peptide ions with $m/z=674.8$). The
intermolecular potential was assumed to be a variable hard sphere for
air species and hard sphere for air-ion collisions.  The reference
diameter and temperature exponent were chosen from Ref.~\citen{DSMC}
for air. Ions had a diameter of 16.82~\AA \cite{Tao}.  Energy
redistribution in molecular collisions between the internal and
translational modes was performed in accordance with the
Larsen-Borgnakke model\cite{LB}.  Temperature-dependent relaxation
numbers were used. The reflection of air molecules on the surface was
assumed to be diffuse with complete energy and momentum accommodation.
A surface temperatures of 300~K was set at the outer capillary
surface, the Einzel lens, and the skimmer.  The surface was assumed
fully adsorptive for ions. The total number of simulated particles was
approximately 50~million with 3~million collision cells. The collision
grid was automatically adapted based on the local density.  The inflow
boundary condition was extracted from the Navier-Stokes field obtained
at the first step, along the $M=3$ isoline. The outflow was a uniform
field with a gas pressure and temperature of 1~Torr and 300~K,
respectively, in the plume expansion chamber, and 0.01~Torr and 300~K,
in the hexapole chamber.

The impact of the electrostatic force on ion motion was taken into
account in the DSMC simulations. The force values were pre-calculated
with the COMSOL software.  The three dimensional geometry used in the
DSMC calculations was imported into COMSOL
Multiphysics\textsuperscript{\textregistered}, in which the
Electrostatics package was used to set potentials on the appropriate
surfaces.  Once set, a three dimensional lookup table was generated
which gave the electric field vectors throughout the DSMC domain.
That lookup table was exported and subsequently used in the DSMC
process to apply spatially appropriate forces on the molecules through
the first order linear interpolation.  Figure~\ref{COMSOL} shows the
individual surfaces used within the COMSOL model and two dimensional
slices of the electric potential and field vectors, respectively, for
a lens potential $U_{lens}=135$~V.

\section{Experimental Setup}

The electrospray source was comprised of a syringe pump and
microfluidic union (IDEX, Oak Harbor, WA) connecting the syringe
needle with a short piece of fused silica capillary terminated with
the ESI tip (New Objective, Woburn MA).  The sprayed acedified (1\%
formic acid) water/acetonitrile solution (70/30\%, v/v) contained
$10^{-5}$~M of peptide substance $P$ ($MW=1347$~Da). All solutions and
peptide were purchased from Sigma (Sigma-Aldrich, St. Louis, MO, USA).
The ion current measurements were performed using a prototype unit
similar to the one used in a portable mass
spectrometer\cite{Misharin}.  The MS inlet was a glass-lined stainless
steel capillary (0.5~mm i.d, 1.6~mm o.d., 10~cm long) from SIS
(Ringoes, NJ) with its head (2~cm long) embedded in a massive copper
block.  The temperature control system maintained a constant
temperature of 120-220\celsius~for the capillary in the subatmospheric
chamber kept at approximately 1~Torr gas pressure.  
A conical skimmer separated the low-vacuum section (1~Torr) of the
interface containing a tube lens and medium-vacuum (0.01~Torr) section
with the hexapole ion guide. Ion current measurements for selected
analyte ions were performed using a quadrupole ion trap.

\section{Gas Flow Through Capillary}

\begin{figure}[htb]
  \begin{subfigure}[b]{.48\linewidth}
    \centering
    \includegraphics[width=.99\textwidth]{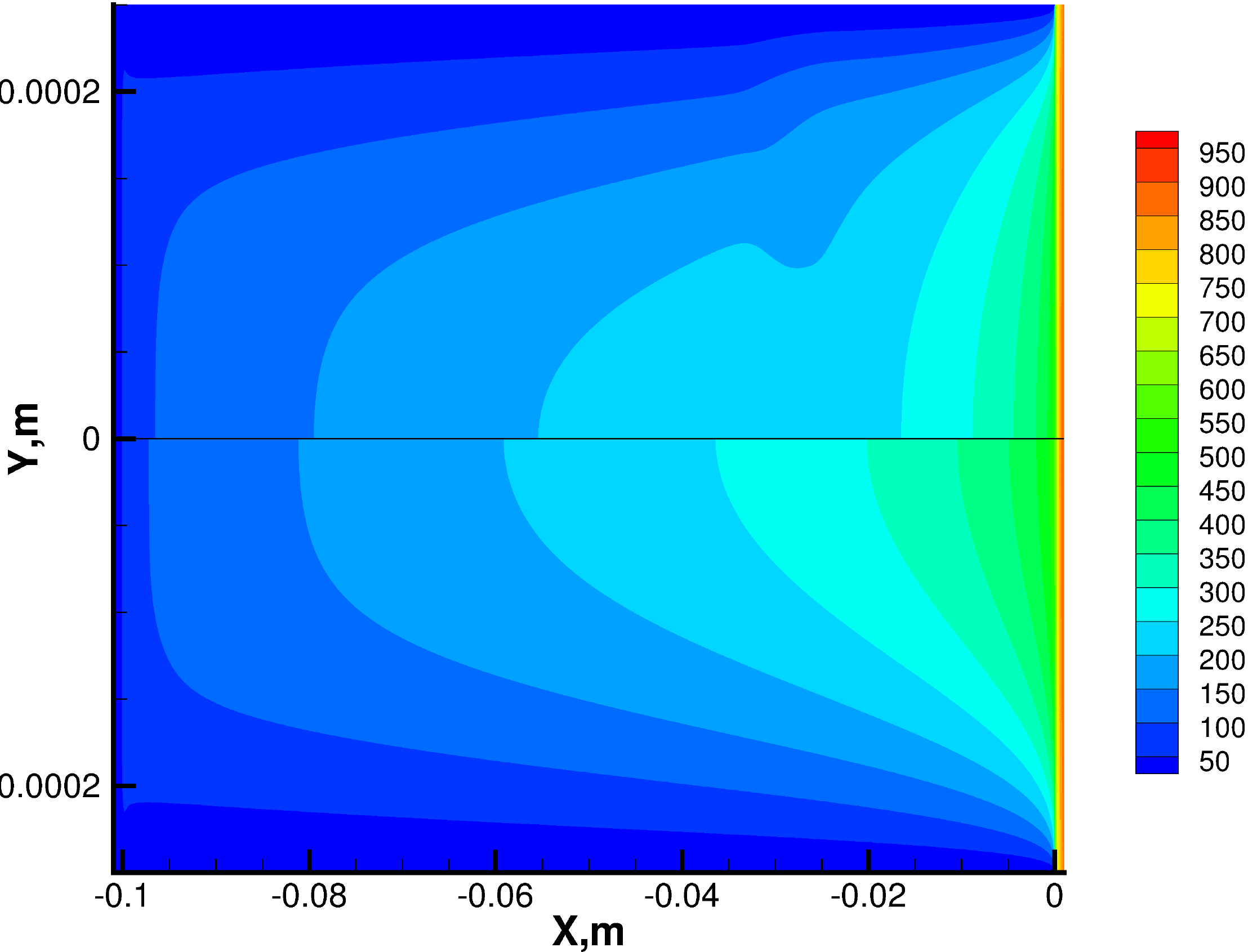}
    \caption{}\label{fig:2a}
  \end{subfigure}%
  \begin{subfigure}[b]{.48\linewidth}
    \centering
    \includegraphics[width=.99\textwidth]{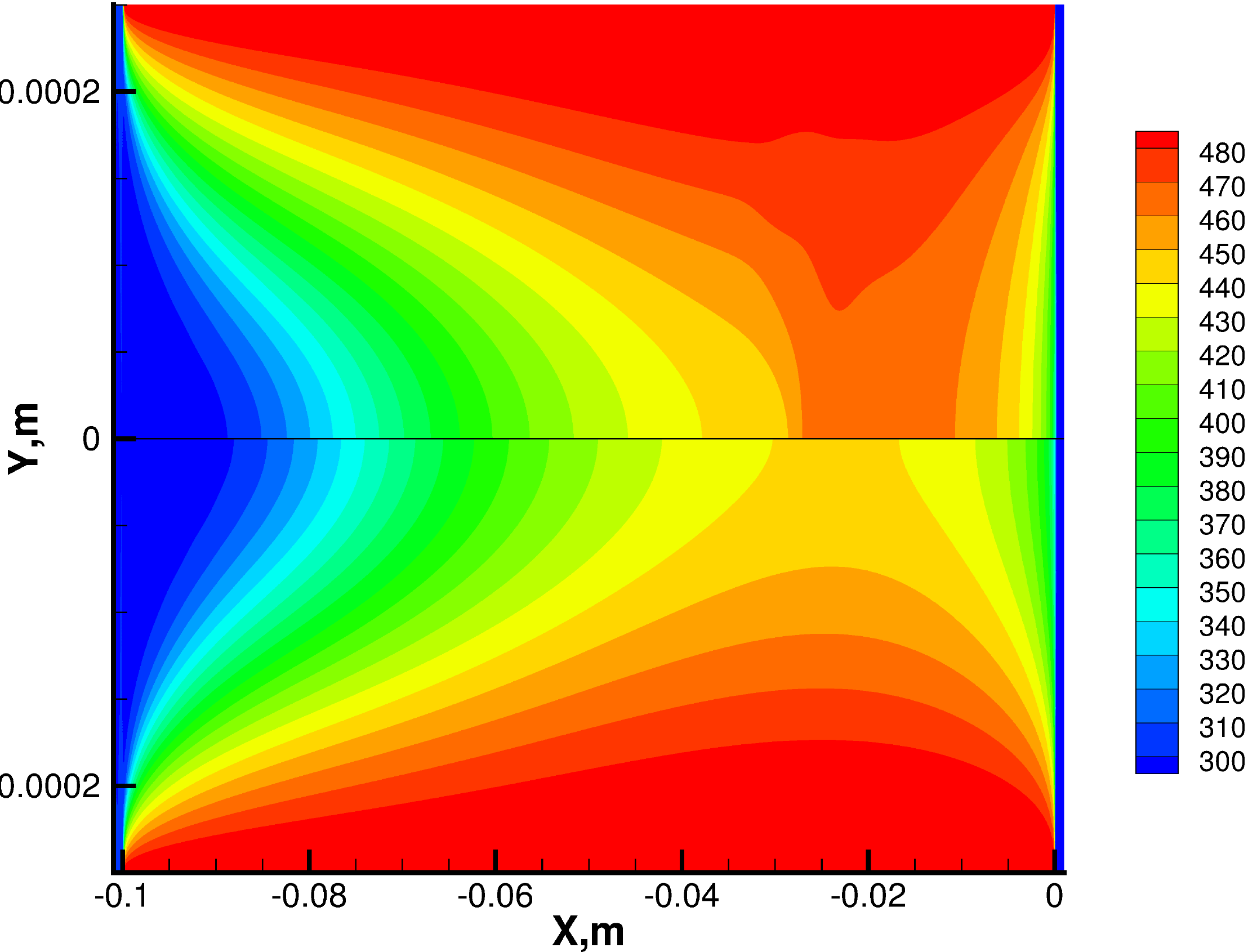}
    \caption{}\label{fig:2b}
  \end{subfigure}%
   \\%
  \begin{subfigure}[b]{.48\linewidth}
    \centering
    \includegraphics[width=.99\textwidth]{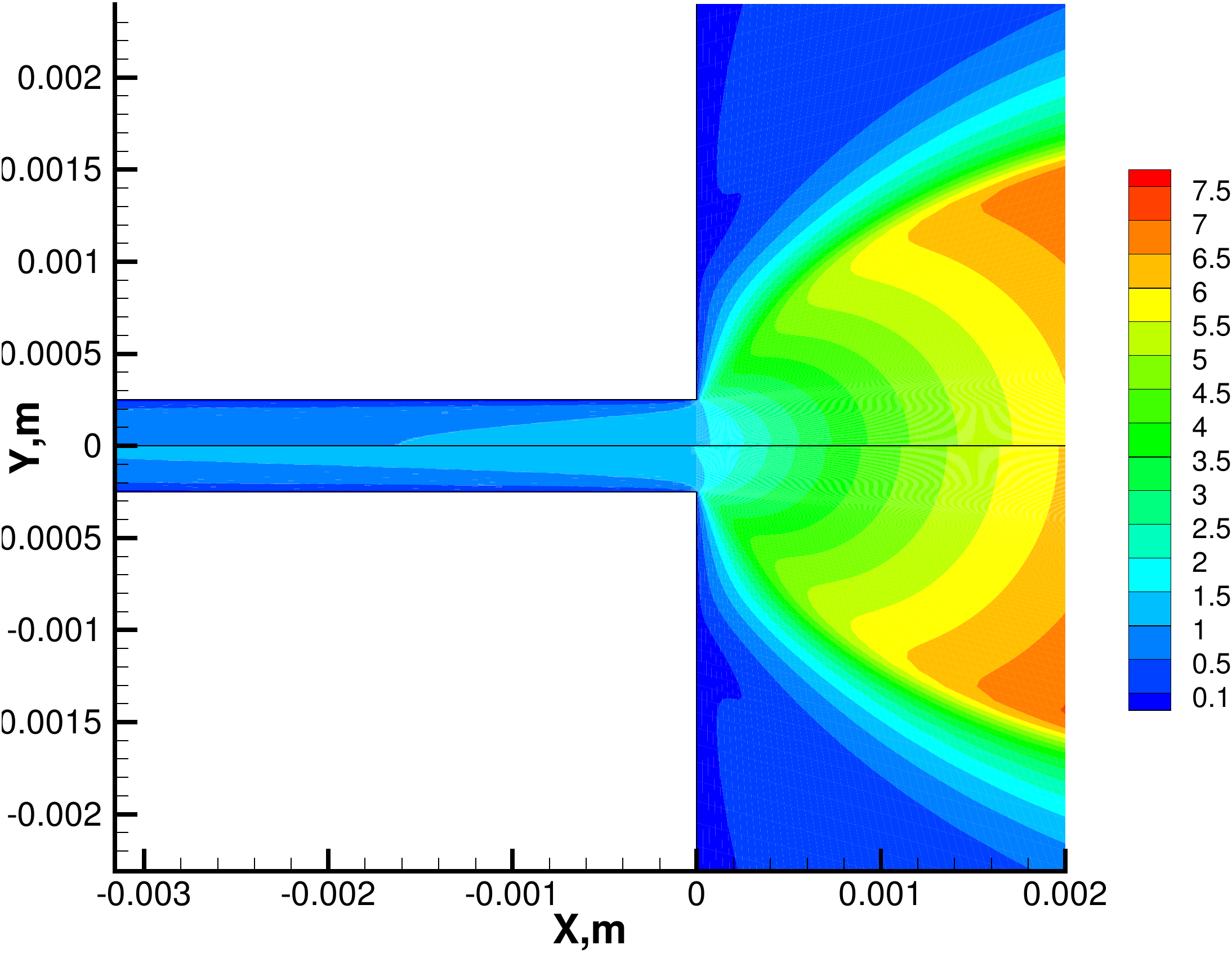}
    \caption{}\label{fig:2c}
  \end{subfigure}%
  \begin{subfigure}[b]{.48\linewidth}
    \centering
    \includegraphics[width=.99\textwidth]{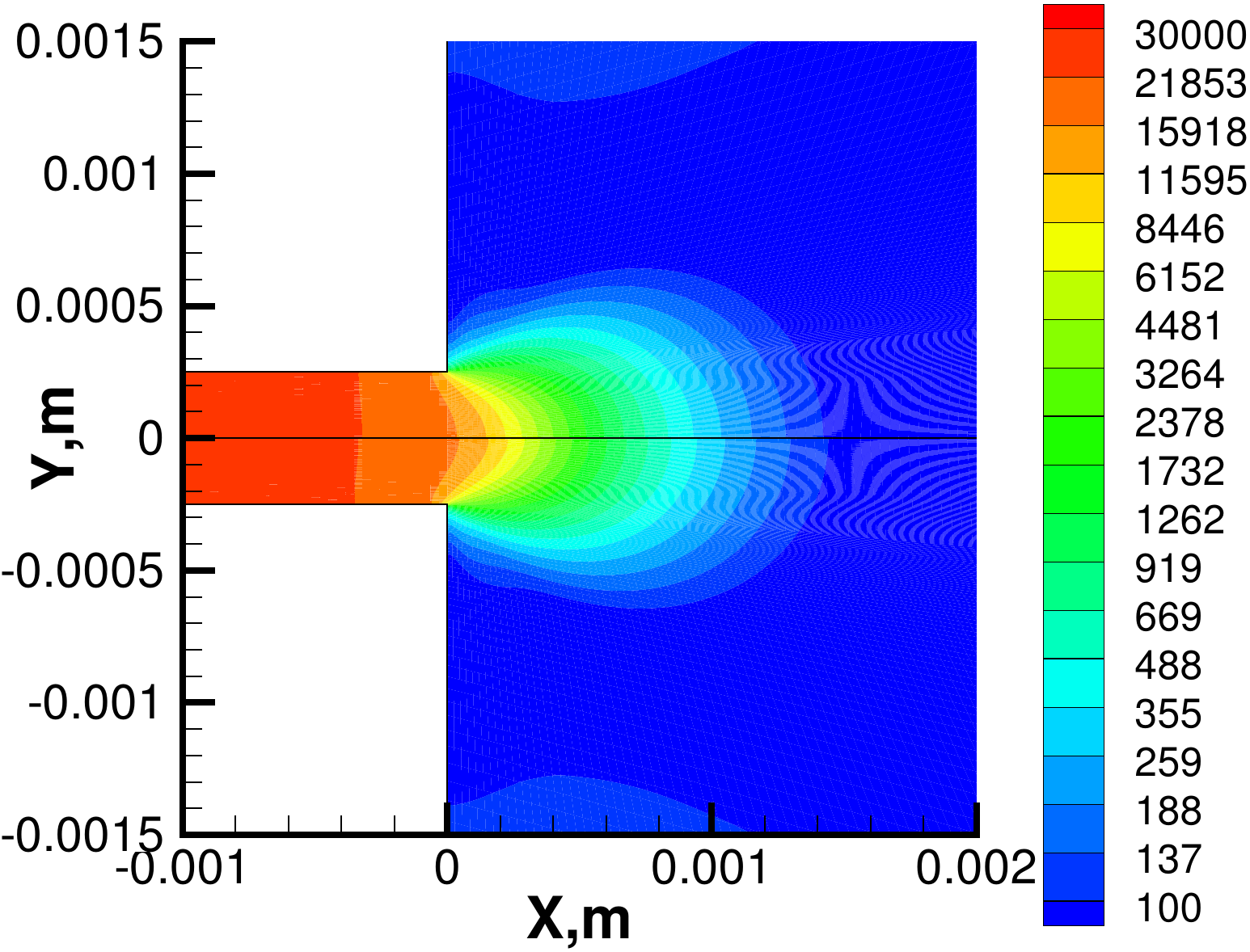}
    \caption{}\label{fig:2d}
  \end{subfigure}%
  \caption{ (a) Carrier gas axial velocity (m/s) and (b) temperature
    (K) inside the capillary. Top halves, turbulent model; bottom
    halves, laminar model. Note X/Y scaling. (c) Carrier Mach number
    and (d) gas pressure (Pa) near the capillary exit. Top halves,
    turbulent model; bottom halves, laminar model.}
\end{figure}

As discussed above, the two-step modeling of ion transport into the
hexapole chamber starts with the Navier-Stokes solution of the
capillary flow. The flow velocity in the axial direction for the
Navier-Stokes computations is shown in Fig.~\ref{fig:2a}. The
Reynolds number in the capillary exceeds 3,000 and thus one can
expect noticeable turbulence effects.  To illustrate,
results are shown for the turbulent and laminar models (the
simulation setup is identical in these computations except for the
turbulence terms being turned on and off).  The result is stretched in Y
direction in order to provide better detail. Most of the ESI and
subatmospheric chambers is not shown. In both cases, the flow velocity
quickly increases downstream of the capillary entrance. The velocity
increase is mostly in the core flow, as the velocity near the wall
does not exceed 10~m/s due to relatively high gas density and
therefore thin boundary layer. As expected, the turbulent flow is
generally characterized by lower velocities, with the difference
between the turbulent and laminar solutions equals to $\sim$~10\% in
most of the capillary. This difference decreases at the capillary
exit, where the maximum flow velocity is about 555~m/s for the laminar
case and 530~m/s for the turbulent case.

The gas temperature for these two cases is shown in Fig.~\ref{fig:2b}.
Note that although the hot wall and gas convection result in a
significant increase in gas temperature in the core flow, it always
stays noticeably lower than the wall temperature. The maximum
temperature at the capillary axis is observed at approximately 80\%
capillary length. Further downstream, flow acceleration due to the
proximity of the exit results in a temperature drop. There is a
visible difference between the turbulent and laminar result, although
most of the difference is observed at the axis, near the maximum
temperature region. The maximum temperature at the axis is 468~K for
the turbulent case and 442~K for the laminar case. This comparison
indicates that turbulent mixing, which results in lower flow velocity
(and thus longer residence time) and higher gas temperature, may be
expected to significantly impact the droplet evaporation and fission
process in the downstream half of the capillary.

Let us now consider the details of the flow in the vicinity of the
capillary exit, as this is where the Navier-Stokes to DSMC hand-off
surface is located. The Mach field is presented in Fig.~\ref{fig:2c}.
Typical for flows in capillaries that connect high and low pressure
chambers, the supersonic flow is observed noticeably upstream of the
capillary exit (which illustrates the limitations of approximations
used in Refs.~\citen{greek,ouyang}). The location of the M=1 isoline
visibly changes when turbulence is accounted for, as it crosses the
flow axis at 4~mm upstream of the exit in the laminar case and at
1.5~mm in the turbulent case. Nevertheless, the difference between the
two cases in the expanding plume is much smaller. In the plume, the
turbulent-vs-laminar difference is only on the order of 3\%, which is
very similar to the 3\% mass flow rates for these cases, 1.25$\times
10^{-5}$~kg/s for turbulent and 1.28$\times10^{-5}$~kg/s for laminar
flow. Generally, these calculations show that while there is a visible
impact of flow turbulence on gas properties inside the capillary, its
impact on the plume, and thus gas properties at the starting surface
located along the M=3 isoline, is minor.

This is further illustrated in Fig.~\ref{fig:2d} where the gas pressure
field is presented. For both cases, pressure gradients inside the
capillary are negligible in the radial direction, except in the
immediate vicinity of the exit. In the plume, the difference between
the turbulent and laminar cases is also negligible. Still, in all
results shown in the following sections, the hand-off surface
extracted from the turbulent flow solution was used. Note that the
pressure at the exit is about 16~kPa, over ten times higher than the
background pressure. The gas mean free path at the exit is about
0.5~$\mu$m, so that the capillary diameter based Knudsen number is
about 0.001, and thus well within the applicability of the continuum
approach (Navier-Stokes). At $M=3$ (location of the hand-off surface),
the mean free path increases only an order of magnitude, so that the
continuum approach may still be used.

\section{Expanding Flow Modeling} 

The information on gas macroparameters in the plume nearfield,
obtained as the result of the capillary flow modeling discussed in the
previous section, allows one to use these macroparameters as the
inflow boundary condition (hand-off surface) in the subsequent DSMC
modeling. The Maxwellian distribution function was assumed for gas
molecules and ions originating at the inflow boundaries. A constant
value of the ion mole fraction of 10$^{-5}$ was used, small enough to
avoid impact on the carrier gas flow.  The ion properties therefore
were assumed to be equal to those of the gas. Note that the gas
density in that region is still too high for noticeable ion-gas
separation, confirmed by comparison of the obtained results with those
for the handoff-surface located at $M=2$ (not shown here).

\begin{figure}[htb]
  \begin{subfigure}[b]{.48\linewidth}
    \centering
    \includegraphics[width=.99\textwidth]{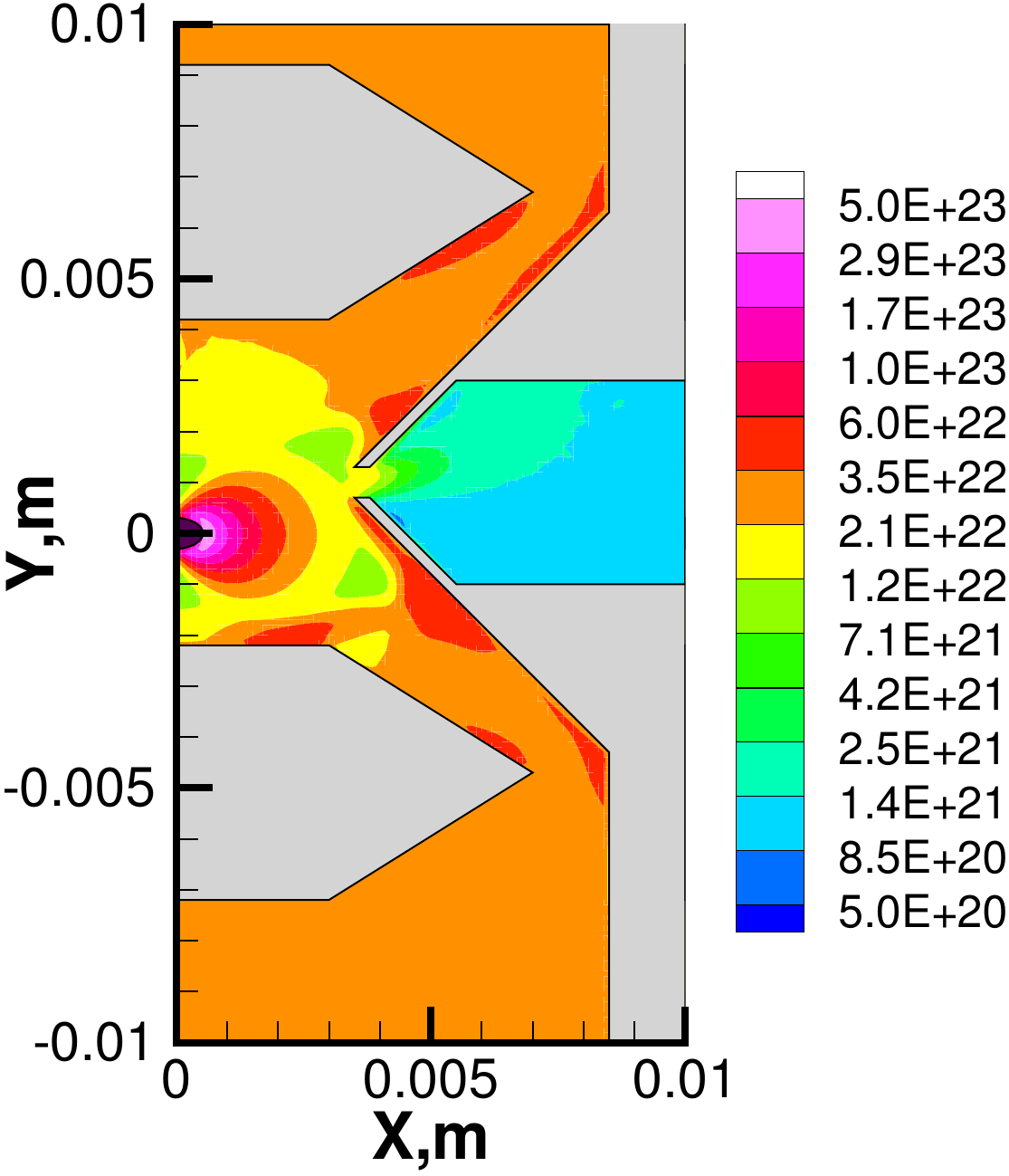}
    \caption{}\label{fig:3a}
  \end{subfigure}%
  \begin{subfigure}[b]{.48\linewidth}
    \centering
    \includegraphics[width=.92\textwidth]{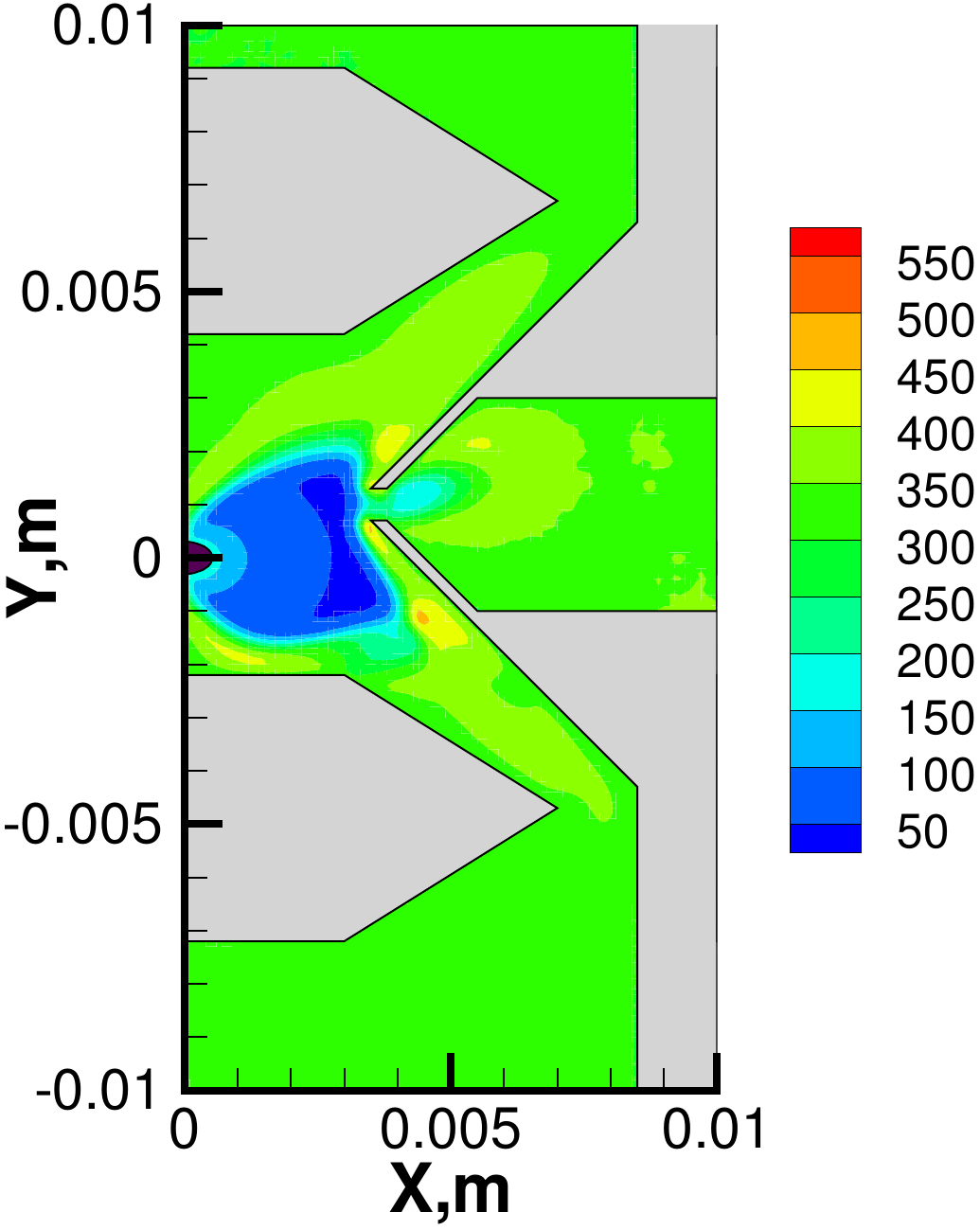}
    \caption{}\label{fig:3b}
  \end{subfigure}%
  \caption{Carrier gas number density (molecule/m$^3$) (a) and
    temperature (b). }
\label{fig3}
\end{figure}

The carrier gas number density in the subatmospheric chamber is given
in Fig.~\ref{fig:3a} for an Einzel lens potential $U_{lens}=0$~V.
Hereafter, a two-dimensional slice of the three-dimensional flowfield
is taken at $Z=0$. The capillary axis is $Y=0$ and its exit plane is
at $X=0$.  The dark region near the origin of the coordinate system
indicates the location of the hand-off surface. The upper and lower
pentagons show the position of the Einzel lens (shifted 1~mm upward in
the $Y=0$ plane), and the remaining surface indicates the location of
the skimmer (centered at $Y=1$~mm).  Although the case of zero Einzel
lens potential is shown here, varying Einzel lens potential does not
impact the carrier gas properties due to the low concentration of
ions. The presented field shows that the plume density reaches the
background gas level after the first two capillary diameters. Further
expansion decreases density down to about 0.5 of that of the
background. Elevated density near the entrance of the skimmer shows
some moderate impact of the plume in that region.  Note also that
plume formation inside the skimmer clearly points upward, which is
also due to the presence of the capillary expansion. It is important
to note that although the low gas density behind the skimmer results
in some plume molecules (and thus ions) being transported through the
skimmer, the vast majority of plume molecules (over 98\%) miss it.
Since the path of ion expansion is not expected to differ strongly
from that of air, such a low transmission provides a clear indication
of the importance of the Einzel lens in the considered setup.

The gas temperature field for $U_{lens}=0$~V is presented in
Fig.~\ref{fig:3b}. In the plume expansion, the gas cools down and the
temperature drops below 50~K (the lowest observed value is 43~K).
The cold region propagates down to the skimmer, almost reaching its
surface. The elevated temperatures near the skimmer (in excess of
400~K) provide some insight into flow non-equilibrium. The reason for
the temperature to be higher than the skimmer surface temperature of
300~K is that gas rarefaction results in a non-Maxwellian, bimodal,
velocity distribution function, which distorts the traditional and
calculated idea of temperature. The bimodality is due to nearly
counter-propagating plume molecules and molecules reflected on the
skimmer surface.  At the entrance of the skimmer, the temperature
reaches about 300~K and then falls again due to gas expansion. The
results show that due to the combination of low temperatures between
the capillary exit and the skimmer, and relatively low densities in
that region, no noticeable change in droplet evaporation due to the
carrier gas will occur downstream of the capillary exit. The
conclusion should still be valid even if the skimmer is heated.

\begin{figure}[htb]
  \begin{subfigure}[b]{.48\linewidth}
    \centering
    \includegraphics[width=.99\textwidth]{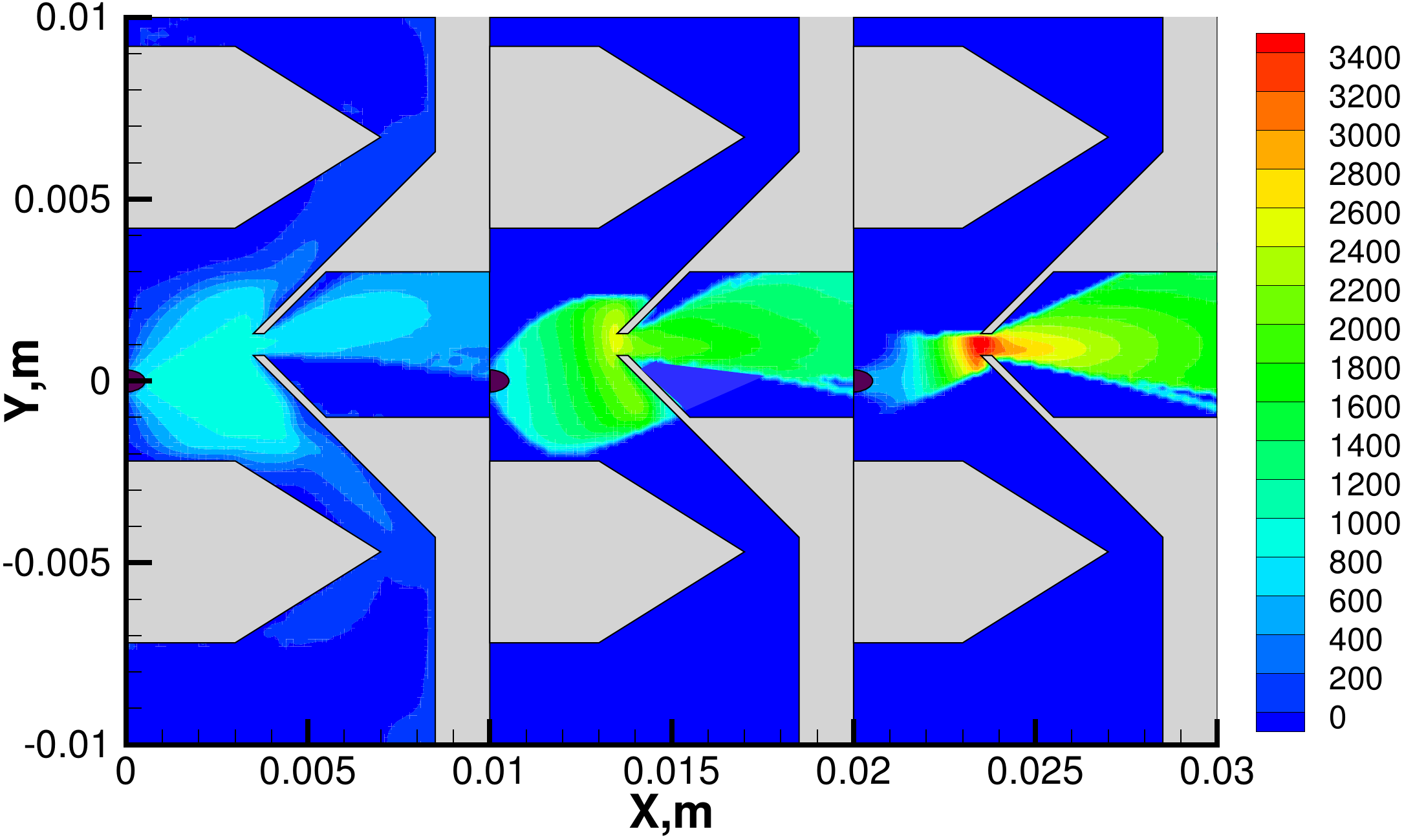}
    \caption{}\label{fig:4a}
  \end{subfigure}%
  \\%
  \begin{subfigure}[b]{.48\linewidth}
    \centering
    \includegraphics[width=.99\textwidth]{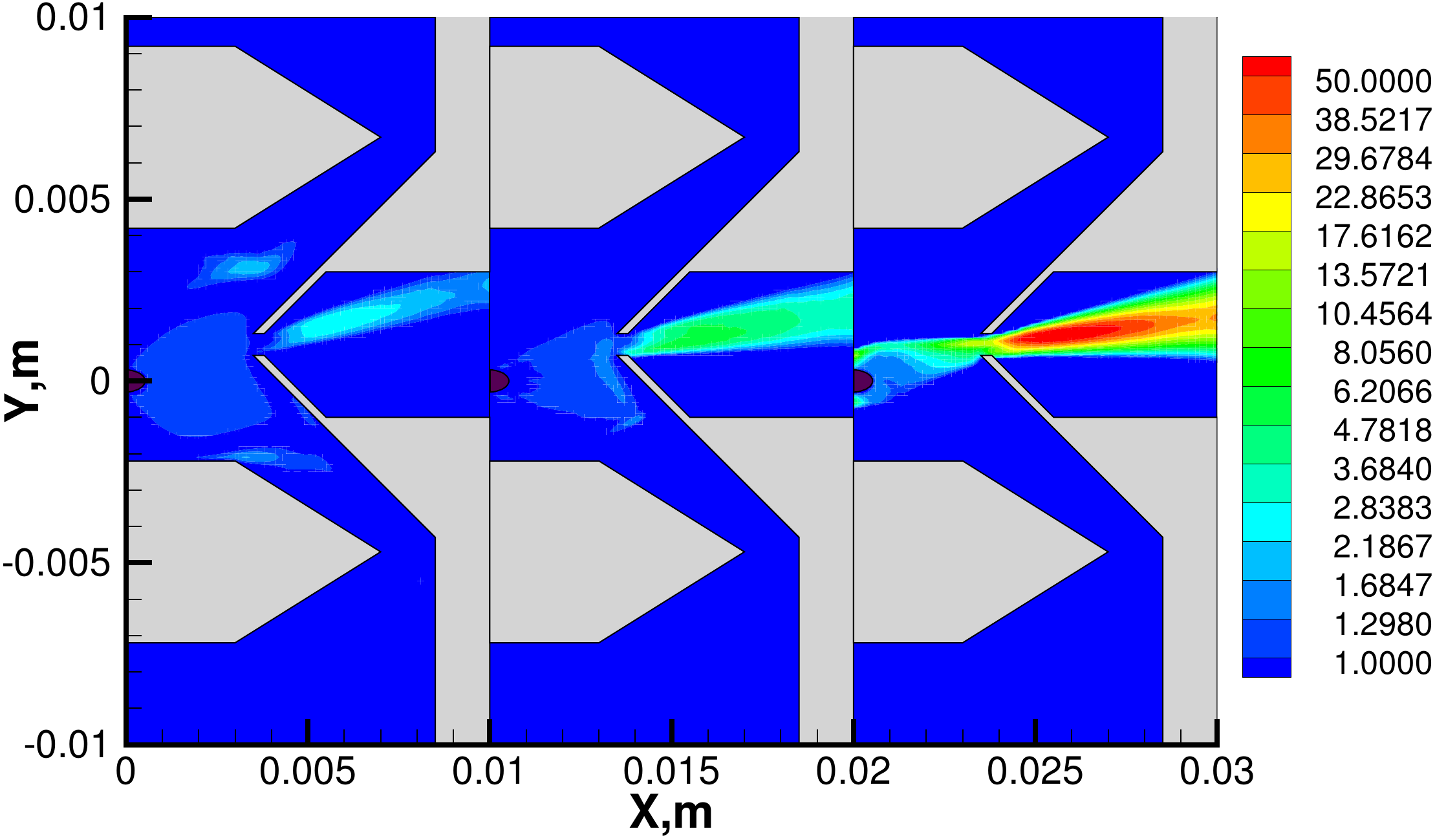}
    \caption{}\label{fig:4b}
  \end{subfigure}%
  \\%
  \begin{subfigure}[b]{.48\linewidth}
    \centering
    \includegraphics[width=.99\textwidth]{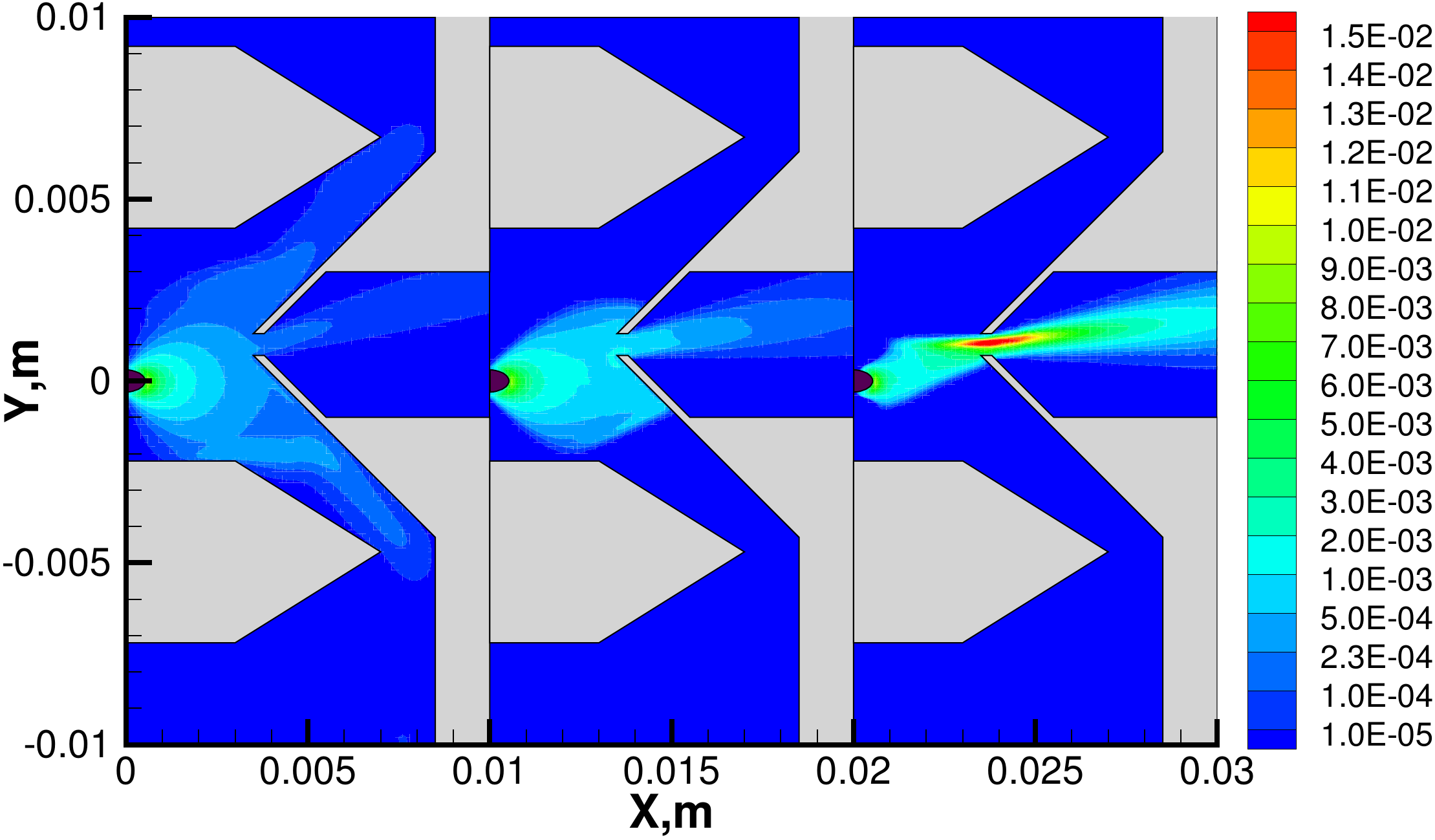}
    \caption{}\label{fig:4c}
  \end{subfigure}%
  \caption{(a) Ion axial velocity $V_x$ (m/s), (b) ion mole fraction
    normalized to its initial value, and (c) ion mass flux (kg m2/s),
    for different Einzel lens potentials. Left, 0~V; center, 45~V,
    right, 135~V.}
\label{fig7}
\end{figure}

Consider now the properties of ions as they are impacted by the Einzel
lens potential. The ion velocity is shown in Fig.~\ref{fig:4a} for the
three values of the Einzel lens potential $U_{lens}$. In the absence of an
applied potential, $U_{lens}=0$, the ion velocities and trajectories largely
follow those of the carrier gas. The maximum flow velocity of ions,
about 930~m/s, is close to that of gas, 960~m/s, which corresponds to
the conventional gas expansion at a stagnation temperature of 300~K.
The maximum velocity is observed at the capillary axis. Some decrease
in ion velocity near the skimmer walls is related to collisions
with gas molecules reflected on the walls. A higher rate of velocity
decrease corresponds to higher gas density near the wall (see
Fig.~\ref{fig3}). Collisions with gas molecules also cause some
decrease in ion velocity downstream of the skimmer entrance. The ion
flow velocity field drastically changes when a 45~V potential is
applied to the Einzel lens. Most importantly, the electrostatic field
created by the capillary-lens-skimmer potential difference starts to
focus the ions to the skimmer. The electrostatic force, on average,
reduces the radial velocity component of ions, which in turn deflects
their trajectories toward the lens axis. In addition, the axial
velocity component noticeably increases as ions travel from the
capillary to the skimmer, reaching about 2,400~m/s near the skimmer
entrance. Collisions with air molecules decrease that velocity to
less than 1,400~m/s a few millimeters downstream of the entrance.  For
a 135~V potential, the focusing effect is even more substantial.
However, in the first two millimeters of the capillary exit, the
average ion velocity is smaller than that for $U_{lens}=45$~V and even
$U_{lens}=0$ case. The reason for this is that the high voltage at the
lens creates an electrostatic field that decreases ion velocity in
that region and pushes some ions back towards the capillary. As the
ions travel toward the skimmer, they are rapidly accelerated to
average velocities of about 3,500~m/s. Then, they are decelerated
inside the skimmer, where the effect of the electrostatic field is
smaller than that of the gas-ion drag. It may be expected that,
similar to ion deceleration, small droplets would also be decelerated,
thus significantly increasing their internal temperatures and
evaporation rates.

When the background pressure in all subatmospheric sections is fixed,
the efficiency of the lens-skimmer setup directly depends on the
amount of ions that come through the skimmer, as well as the amount of
air molecules from the plume that the skimmer prevents from entering
into the hexapole chamber. A good indicator of the setup efficiency is
therefore the ion mole fraction, with higher mole fractions behind the
skimmer being desirable. The ion mole fraction fields for three
different lens voltage are given in Fig.~\ref{fig:4b}.  All mole
fractions were normalized by the value set at the inflow boundary to
provide a more informative comparison. For the $U_{lens}=0$~V, as can
be expected for a heavy-light gas mixture expansion, the ions are
concentrated near the capillary axis, although their normalized mole
fraction does not exceed 1.1. The ion concentration decreases for
large angles from axis, approaching 0 for 90$^\circ$. The largest ion
mole fraction is observed inside the skimmer, where it reaches 2.8 due
to smaller deflection of ions from their original plume trajectories,
as compared to air molecules. This trend continues for
$U_{lens}=45$~V, although in the latter case the impact of the
electrostatic force from the Einzel lens results in maximum ion mole
fraction of more than 5 that of the inlet. Moreover, the lens
influences the average angle of the ions that come through the
skimmer, turning the ions toward the skimmer axis. Note that both for
$U_{lens}=0$ and $U_{lens}=45$~V, the mole fraction of ions drops near
the skimmer walls. This is due to the fully adsorptive wall condition
assumed for ions in this work. Note that an incomplete adsorption is
not expected to change the flow inside the skimmer, as the vast
majority of reflected ions would be pumped out of the chamber. For the
135~V, case, one can clearly see the backflow of ions near the
starting surface that are diverted by the field toward the capillary
surface. Of the ions that are not diverted, most travel through the
skimmer entrance, where the resultant normalized ion mole fraction
reaches 70, or 25 times that for $U_{lens}=0$~V. Some decrease of the
mole fraction further downstream is due to accumulation of air
molecules reflected on the horizontal skimmer walls.

The most important flow property, from the standpoint of system
efficiency is the ion mass flux, which linearly impacts the MS
reading. The impact of the lens potential on the ion mass flux is
presented in Fig.~\ref{fig:4c}. In the absence of the lens
potential the ion mass flux is maximum at the capillary axis and
decreases in radial direction nearly following the cosine law.  Some
deviation from the cosine law at larger distances from the axis is
related to the gas density variation in those regions. For the
$U_{lens}=45$~V case, ion focusing by the lens is obvious and the
result is a factor of five increase in the ion mass flux at the
skimmer entrance as compared to $U_{lens}=0$. Also, although none of the
ions are diverted by the lens back to the capillary wall, a large
amount of ions end up on the skimmer walls. For both 0 and 45~V, the
variation of the ion mass flux across the skimmer entrance is small.
For the 135~V potential, that variation is over three times and the
maximum value is over 15 times larger than that for 45~V (it is in
fact even larger than that inside the capillary). Note that most of
the ions that approach the skimmer come through its entrance and only
few stick to the wall (all of them in the lower half of the skimmer).
This may be a clear indication that for a given geometrical setup
(skimmer and Einzel lens size and location), further increase in 
lens voltage may not increase the ion mass flow through the capillary.
This is because the ion back flow will further increase with the
voltage, while there is a limited amount of ions that can be better
focused into the skimmer.

\begin{figure}[htb]
  \includegraphics[width=11cm]{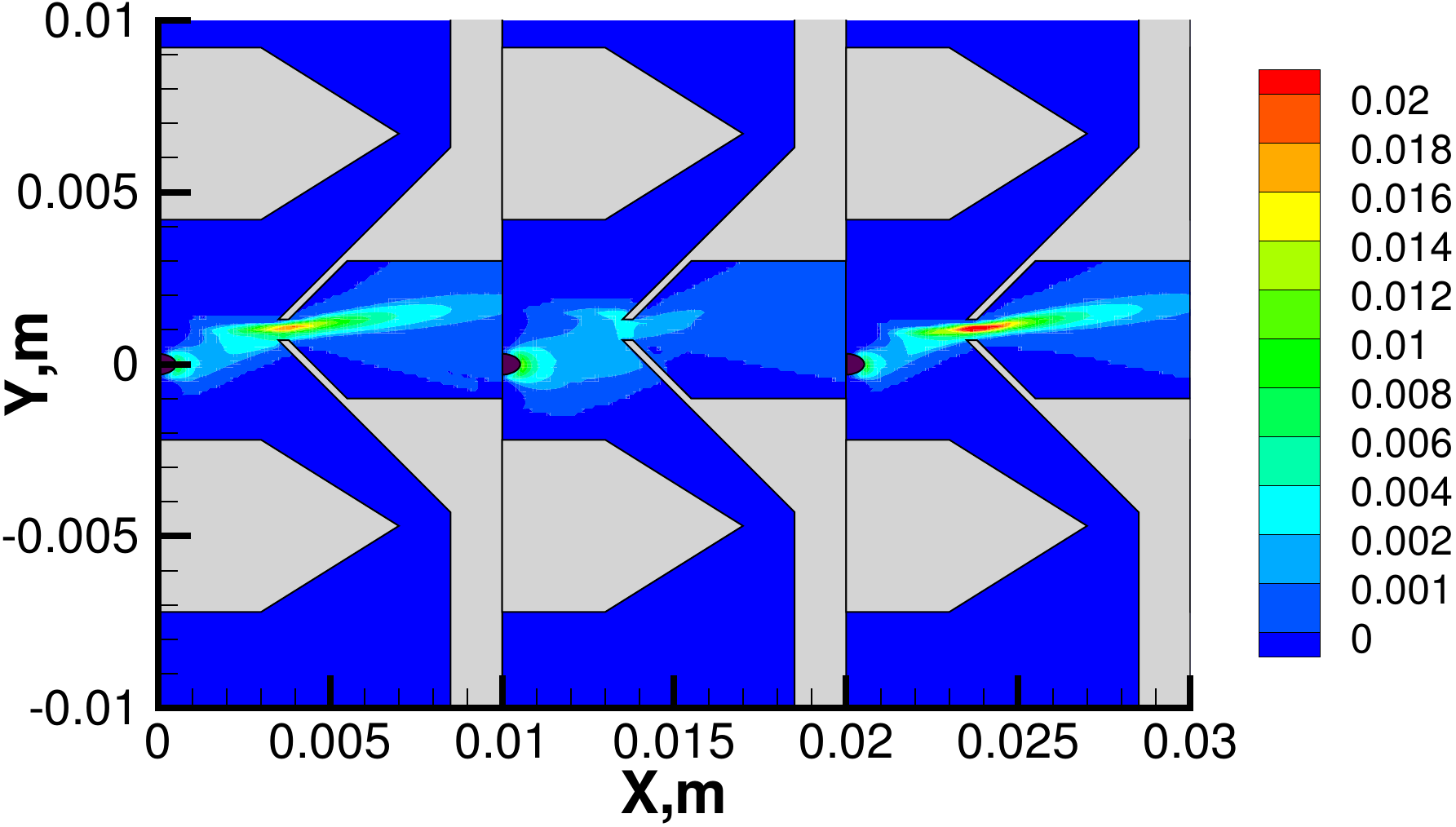}
  \caption{Ion mass flow (kg~m$^2$/s) calculated for ions with
    different $m/z$ or gas/ion collision diameters.  From left to
    right: model ions ($m/z=674.8$, $z=+2$); singly-protonated water
    clusters (100-mers, $m/z=1802.1$); model ions ($m/z=674.8$,
    $z=+2$) but with 20\% reduced gas/ion collision diameter.  }
\label{fig:5}
\end{figure}

Prior to comparing the numerical results to the experiment, it is
reasonable to study the sensitivity of such a comparison to some
numerical and experimental uncertainties. The carrier gas related
uncertainty is considered relatively minor. The focus below is on ion
related changes. The two points examined here are the impact of the
uncertainty in the ion diameter, which directly influences the ion-gas
interaction, and possible MS signal contamination by water droplets.
Figure~\ref{fig:5} compares the ion mass fluxes, normalized by the
inflow values, obtained for the following computations, all conducted
at $U_{lens}=135$~V: (i) the model ions used above (i.e., peptide ions
with $m/z=674.8$, doubly charged, with the ion/gas collision diameter of
16.82~\AA), (ii) water clusters (100-mers) carrying a single charge,
and (iii) peptide ions with $m/z=674.8$, doubly charged, but with a
diameter reduced to 14~\AA.  The results show that for the same
concentration of charged water clusters, their mass flux is noticeably
different.  The lower charge-to-mass ratio results in poor focusing by
the lens, and many droplets will collide with the skimmer surface,
thus reducing contamination. The mass flux at the skimmer entrance is
on average approximately a factor of 3.5 smaller than for the baseline
model.  Note that the mass flux at the entrance should decrease with
the increase of the droplet mass. Direct analysis of contamination of
the signal with charged water clusters is complicated by possible
droplet heating and evaporation downstream of the skimmer entrance due
to droplet deceleration by the air flow, but it is believed that such
contamination does not have determining effect on the signal.
Although the ion diameter used above is believed to be accurate within
a few percent, a 20\% smaller diameter is used in order to estimate
its possible impact, even though such a value is extremely unlikely.
The decrease of the effective peptide ion/gas collision diameter
results in somewhat better focusing into the skimmer entrance, as the
maximum value of the mass flux is over 40\% larger than that for the
same ion but with ion/gas collision diameter of 16.7\AA. On the other
hand, smaller drag results in more ions being pushed back to the
capillary by the electric field, so that the total mass flow through
the skimmer is in fact 10\% smaller than that for the ions with larger
collision diameter ions.  Generally, the impact of the ion
diameter and ion-air collision uncertainty is believed to be small.

\section{Comparison with Experimental Data}

The Einzel lens voltage was varied, both numerically and
experimentally, and the output ion signal recorded. While the
experimentally measured signal directly reflects the number of doubly
charged peptide ions trapped into the MS, the numerical signal is
based on the number flux of ions that reached the internal part of the
skimmer. The possible ion loss further downstream is not taken into
consideration.  Although the impact of such an approximation is small,
it may still somewhat reduce the high-to-low voltage signal ratio, as
the ion flow passing through the skimmer entrance is more divergent
for lower $U_{lens}$ (see Fig.~\ref{fig:4a}).  Another numerical
uncertainty is related to the use of a two-step approach, which is
also expected to skew the signal ratio toward the lower voltage as
compared to a hypothetical approach that would compute all droplet/ion
evolution in one step. This is because the two-step approach does not
track ion trajectories before they reach $M=3$ isoline, assuming that
they travel with the flow and neglecting the electrostatic field
effect. The numerical uncertainties related to other factors, both
physical (viscosity, heat conductivity, etc.) and numerical (numbers
of cells, simulated particles, timestep, etc.), as well as the
statistical error, are believed to be small (less than 3 percent
combined).

\begin{figure}[htb]
  \includegraphics[width=8cm]{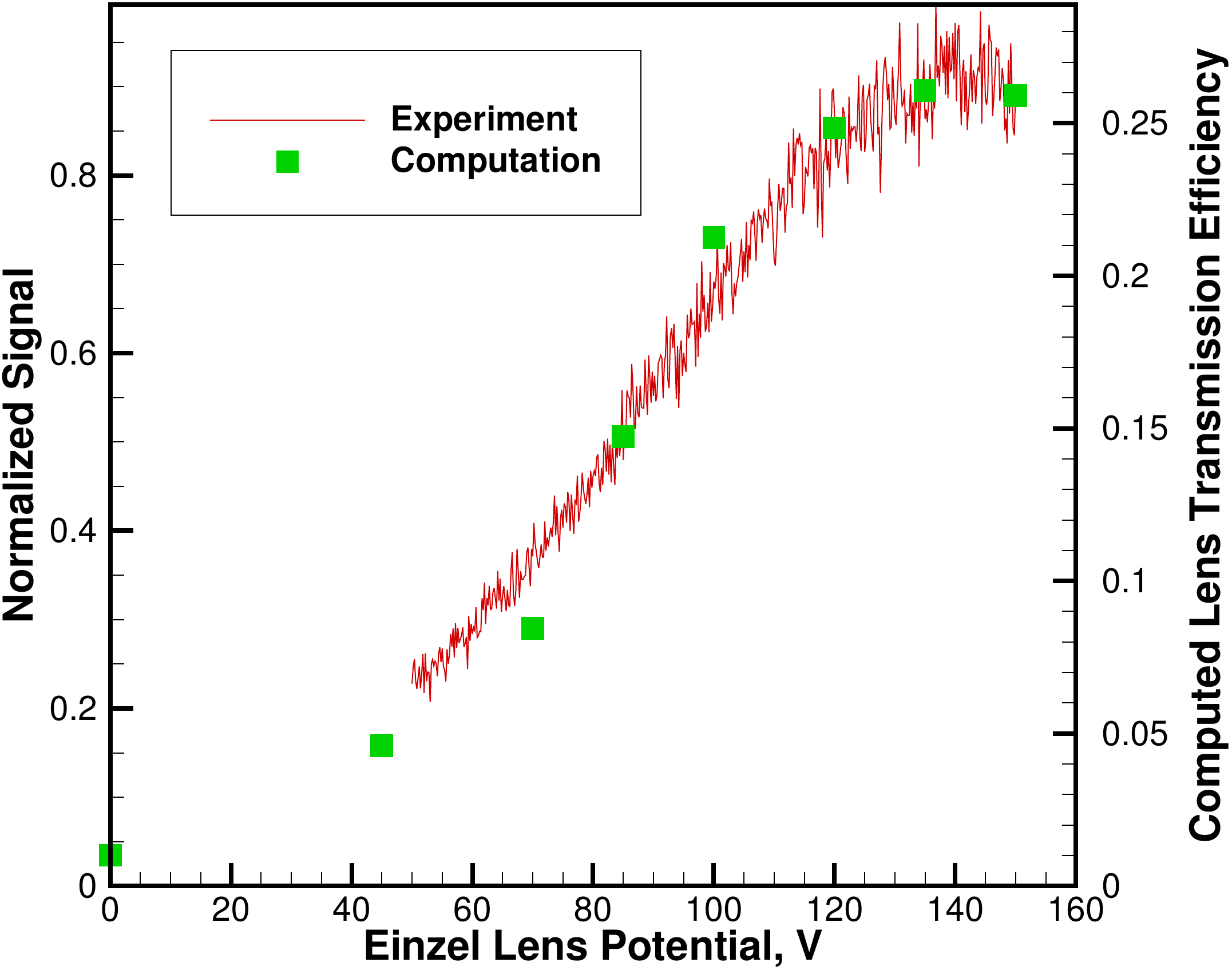}
  \caption{Comparison of measured and computed normalized signal as a
    function of Einzel lens potential. }
\label{fig:6}
\end{figure}

The comparison of measured and computed ion signal is shown in
Fig.~\ref{fig:6}, where the experimental data is normalized by the
maximum signal, and the numerical points are normalized to provide the
best fit with the data. Numerical modeling captures the signal
increase with voltage quite reasonably. Both computation and
experiment show the maximum signal at about 135~V, after which the
signal starts to drop due to ion diversion back to the capillary
surface.  The underprediction of ion signal in the computations for
smaller voltages (less than 100~V) is attributed primarily to the
effects of the signal calculation and two-step approach, discussed
earlier in this sections. Figure~\ref{fig:6} also presents the Einzel
length efficiency (the right axis), calculated numerically as the
ratio of ions passed through the skimmer to the number of ions that
left the capillary. It is seen that although the lens allows for great
increase in transmission, from less than 2\% for $U_{lens}=0$ to over
25\% for $U_{lens}=135$~V, this experimental configuration still shows
large room for improvement. Even for $U_{lens}=135$~V, almost 75\% of
ions are lost at the skimmer and external capillary walls.

\section{Conclusions}

A two-step numerical approach for ion transport from a capillary to a
skimmer in a conventional ESI-MS setup is presented. The approach is
based on the solution of the Navier-Stokes equations in the capillary
and the solution of the Boltzmann equation for the gas/ion transport
from the capillary to the skimmer. The direct simulation Monte Carlo
method is used to solve the Boltzmann equation, with a pre-calculated
electrostatic field force term defined by the potential difference
between the capillary and the skimmer. The solvers used in this work
are CFD++ for the Navier-Stokes equations, SMILE for the Boltzmann
equation, and COMSOL for the electric field. The two-step approach is
applied to predict the change in the peptide ion signal as a function
of the Einzel lens voltage. A companion experimental study was also
performed. Good agreement between the measured and the computed ion
signals was observed.  An optimization of the potential applied to the
lens resulted in approximately 10-fold increase of ion transmission as
compared to the case where no focusing field was present. The increase
is attributed to efficient focusing of ions accelerated to the skimmer
entrance.  Still, even for the optimum voltage, maximum
capillary-to-skimmer ion transmission was found to be only around
25\%, with the remaining 75\% of ions being lost on the external
capillary surface and, to a lesser degree, on the skimmer surface. Ion
loss near the skimmer due to peptide ion overheating and following
fragmentation was not considered in this model, yet, no in-source
fragmentation was observed in the companion experiment.  Obviously,
the overheating may be an issue for more fragile ions.

Gas flow inside the considered 100~mm long, 0.5~mm diameter, capillary
is turbulent ($Re\approx 3,000$), and the analysis of the turbulence
effect has been conducted.  It was found that flow turbulence
noticeably decreases the flow velocity and increases the maximum
temperature in the core flow and thus will affect the droplet
evaporation process.  The presence of turbulence has only small effect
on the gas flow properties downstream of the the exit of the heated
capillary.  The temperature in the expanding plume stays below 100~K
almost to the surface of the skimmer.  Nanodroplets and water clusters
which can reach that region are not expected to evaporate in
collisions with residual gas, even if one uses elevated temperatures
for the inlet capillary and skimmer. In the presence of significant
potential at the Einzel lens, the ions were found to accelerate to
over 3,000 m/s upstream of the skimmer orifice, and then slowly
decelerate through ion-air collisions downstream of the skimmer
orifice. Such a deceleration leads to heating and evaporating of small
droplets or clusters traveling through the skimmer and to collisional
activation of ion vibrational modes (skimmer-induced fragmentation).

\section{Acknowledgments}

The authors wish to thank Prof. Rebecca Webb and her students for
their helpful discussion regarding COMSOL.  The work was supported in
part by the Air Force Office of Scientific Research (Dr. Mitat
Birkan). The modeling effort used, in part, the Extreme Science and
Engineering Discovery Environment (XSEDE), which is supported by
National Science Foundation grant number OCI-1053575.  This work was
also supported, in part, by a grant of computer time from the DoD HPC
Modernization Program (HPCMP) at the ERDC DoD Supercomputing Resource
Center (DSRC). The experimental efforts have been supported in part by
NIH Grant 1R43GM103358.

\end{document}